\documentclass[preprint,aps,showpacs]{revtex4}
\usepackage{graphicx,epsf}
\usepackage{lscape}
\usepackage{citesort}
\begin{document}

\title{{\Large{\bf Exclusive
 $D_{s} \to (\eta,\eta^{\prime}) l \nu$
decays in light cone QCD}}}

\author{\small
\small K. Azizi$^1$ \footnote{e-mail: kazizi @ dogus.edu.tr}, R.
Khosravi$^2$ \footnote {e-mail: khosravi.reza @ gmail.com}, \small
F. Falahati$^3$ \footnote {e-mail: falahati@shirazu.ac.ir}}

\affiliation{$^1$Physics Division, Faculty of Arts and Sciences,
Do\u gu\c s University, Ac\i badem-Kad\i k$\ddot{o}$y, 34722
Istanbul, Turkey \\$^2$Physics Department, Jahrom Higher Education Complex, 74137 Jahrom, Iran  
\\$^3$Physics Department, Shiraz University,
Shiraz 71454, Iran}

\begin{abstract}
Probing the $\bar ss$ content of the $\eta$ and $\eta'$ mesons and considering mixing between these states as well as  gluonic contributions,  the form factors responsible for
 semileptonic $D_s \to (\eta, \eta') l \nu$ transitions  are
calculated via  light cone QCD sum rules. Corresponding branching fractions and their ratio for different mixing angles are also obtained.  Our results
are in a good consistency  with experimental data as well as
predictions of  other nonperturbative approaches.
\end{abstract}

\pacs{11.55.Hx, 13.20.-v, 13.20.Fc}

\maketitle

\section{Introduction}
Based on experimental results, a considerable part of the total decay rate of the $D_s$ meson is related to its decay to $\eta$ and $\eta'$ mesons. Therefore, the $D_s$ is a proper meson to study the phenomenology of
the  $\eta$ and $\eta'$ mesons and their structures. Due to charm quark, this meson plays an essential role in  analyzing of the weak and strong interactions as well as exploring new physics beyond the standard model (SM)
 which will be probed by the large hadron collider (LHC). The charmed systems are known for very small CP violations in the SM, hence any detection of CP violations in such systems can be considered as a signal for
 presence of new physics (for more information about the $D_s$ meson and its decays see \cite{Rydpet}).

In the present work, we analyze the semileptonic $D_s \to (\eta,
\eta') l \nu$ decays in the framework of  light cone QCD sum
rules (LCSR). The $\eta$ and $\eta'$ mesons are mixing
states \cite{FKS,DP},
\begin{eqnarray}\label{eq11}
|\eta\rangle &=&  \cos \, \varphi |\eta_q\rangle- {\sin} \, \varphi  |\eta_s\rangle, \nonumber \\
|\eta'\rangle &=& { \sin} \, \varphi  |\eta_q\rangle+{\cos} \,
\varphi |\eta_s\rangle ,
\end{eqnarray}
where $\varphi$ is single mixing angle. The  measured values of $\varphi$ in the the quark flavor (QF) basis (for more information about this basis  see for instance \cite{HMC,CCD,vincent,ek}) are
$\varphi=(39.7\pm 0.7)^{\circ}$ and $(41.5\pm 0.3_{\rm stat}\pm
0.7_{\rm syst}\pm 0.6_{\rm th})^{\circ}$ with and without the
gluonium content for $\eta'$, respectively \cite{KLOE}.  The mixing angle $\varphi$ has also been obtained as $\varphi=[39.9\pm 2.6 (exp)\pm2.3(th)]^{\circ}$ by recently
 measured $BR[D(D_s)\to\eta(\eta')+\bar l+\nu_l]$ in light-front quark model \cite{ke}.

 In QF
basis,
\begin{eqnarray}\label{eq12}
|\eta_q \rangle &=&{1 \over \sqrt{2} } \left( |\bar
{u} u\rangle +|\bar{d} d\rangle\right), \nonumber \\
|\eta_s \rangle &=& |\bar{s} s\rangle \,\, .
\end{eqnarray}
 Since the $D_s$ meson decays to $\eta$ and $\eta'$ via $\eta_s$
state, the transition form factors of these decays in the QF basis
 are written in terms of the
transition form factors of $D_s\to\eta_s$ as:
\begin{eqnarray}\label{eq13}
f_i^{D_s\to\eta} = -\sin\varphi\times f_i^{D_s\to\eta_s} ,\qquad
f_i^{D_s\to\eta'} = \cos\varphi\times f_i^{D_s\to\eta_s}.
\end{eqnarray}
For calculation of $f_i^{D_s\to\eta^{(\prime)}}$ via the LCSR through
$f_i^{D_s\to\eta_s}$, information about distribution amplitudes (DA's) of the $|\eta_s\rangle$ state as well as corresponding parameters are needed. These quantities have not
been  known yet, exactly. However, the same  quantities  for $\eta$
meson are available and investigation of $f_i^{D_s\to\eta}$ is
possible, directly. On the other hand according to Eq. (\ref{eq13}),
there is a relation between $f_i^{D_s\to\eta}$ and $f_i^{D_s\to\eta'}$,
\begin{eqnarray}\label{eq14}
{|f_i^{D_s\to\eta}(q^2)| \over |f_i^{D_s\to\eta'}(q^2)|}=\tan \varphi,
\end{eqnarray}
so, our strategy will be as follow.  First, we will calculate the form factors, $f_i^{D_s\to\eta}$ via the LCSR,  then using Eq.
(\ref{eq14}) and the values of the mixing angle $\varphi$, we will evaluate the
transition form factors of $D_s\to \eta' l \nu$.

The paper is organized as follows. In the next section, we obtain the   LCSR for the transition
form factors responsible for $D_s\to \eta l \nu$ decay. Section III
is devoted to the numerical analysis of the form factors and calculation of
branching ratios of the $D_s\to(\eta, \eta') l \nu$ decays. We also compare the obtained
results with the existing predictions of the other nonperturbative approaches as well as experimental data.

\section{LCSR for $D_s\to\eta$ transition form factors }

To calculate the transition form factors of the $D_s
\to \eta$ in LCSR method, we consider the following
correlation function:
\begin{eqnarray}\label{eq21}
\Pi_\mu(p,q) = i \int d^4 x e^{i q x} \langle \eta(p) | T\left\{
\bar s (x) \gamma_\mu (1-\gamma_5) c(x) \bar c(0) i (1-\gamma_5)
s(0) \right\} | 0 \rangle,
\end{eqnarray}
where we will use the DA's of the $\eta$ meson. The main reason for choosing the Chiral current,
$\bar c i (1-\gamma_5) s$  instead of the usual pseudoscalar (PS), $\bar c i
\gamma_5 s$ is to eliminate effectively the  contribution of the twist-3 wave functions which are poorly known and cause the main uncertainties
to the sum rules. This current provides results with less uncertainties
(see also \cite{Huan,Huan2,tm1,tm2,Chern}). Here, we should stress that the Chiral current may enhance the NLO twist-2 contribution and to get more exact results,
 one should use the DA's of the  $\eta$ mesons up to NLO which are not available yet.

 According to the general philosophy of the QCD sum rules and its extension, light cone sum rules, we  should calculate
the above correlation function in two different ways.
In phenomenological or physical  representation, it is calculated in terms of hadronic parameters. In QCD side, it is obtained  in terms of DA's and QCD degrees of freedom.
 LCSR sum rules for the  physical quantities like form factors are
acquired  equating  coefficient of the sufficient structures from
both representations of the same correlation function through
dispersion relation and applying  Borel transformation and continuum subtraction to
suppress the contributions of the higher states and
continuum.

To obtain the phenomenological representation  of
the correlation function, we insert a complete set
of $D_s$ states between the currents.  Isolating the pole term
of the lowest PS $D_s$ meson, we get,
\begin{eqnarray}\label{eq22}
\Pi_\mu(p,q) &=& \frac{ \langle \eta(p) | \bar s
\gamma_\mu(1-\gamma_5)c | D_s(p+q) \rangle \langle D_s (p+q)| \bar c i
(1-\gamma_5) s | 0 \rangle}{m_{D_s}^2 - (p+q)^2}+\cdots,
\end{eqnarray}
where $\cdots$ stands for  contributions of the higher
states and continuum. The  matrix element, $\langle D_s
|\bar c i(1-\gamma_5)s |0 \rangle$ is defined as:
\begin{eqnarray}\label{eq23}
\langle D_s | \bar c i(1- \gamma_5) s | 0 \rangle =
\frac{m_{D_s}^2 f_{D_s}}{m_c+m_s}\,,
\end{eqnarray}
where $f_{D_s}$ is
leptonic decay constant of $D_s$ meson.
 The  transition matrix element, $\langle \eta(p) | \bar s
\gamma_\mu(1-\gamma_5)c | D_s(p+q) \rangle$  can be parameterized via  Lorentz invariance and parity considerations as \cite{tm1,Huan2}:
\begin{eqnarray}\label{eq24}
\langle \eta(p) | \bar s \gamma_\mu (1 - \gamma_5) c | D_s(p+q)
\rangle = 2 f_+^{D_s\to\eta}(q^2) p_\mu + (f_+^{D_s\to\eta}(q^2) +
f_-^{D_s\to \eta}(q^2)) q_\mu \, ,
\end{eqnarray}
where, $f_\pm^{D_s\to\eta}(q^2)$ are  transition form factors responsible for $D_s\to\eta$ decay.
Using Eqs. (\ref{eq23}) and (\ref{eq24})  in Eq.
(\ref{eq22}), we obtain,
\begin{eqnarray}\label{}
\Pi_\mu(p,q) =\Pi_1(q^2, (p+q)^2) p_\mu + \Pi_2 (q^2, (p+q)^2) q_\mu,
\end{eqnarray}
where,
\begin{eqnarray}\label{eq25}
\Pi_1 &=& \frac{2 f_+^{D_s\to\eta}(q^2) m_{D_s}^2 f_{D_s}}{(m_c+m_s)
(m_{D_s}^2 - (p+q)^2)} + \int_{s_0}^\infty ds
\frac{\rho_1^h(s)}{s-(p+q)^2} + \mbox{subtractions}\, , \nonumber \\
\Pi_2 &=& \frac{(f_+^{D_s\to\eta}(q^2) + f_-^{D_s\to\eta}(q^2))
m_{D_s}^2 f_{D_s}}{(m_c+m_s) (m_{D_s}^2 - (p+q)^2)} +
\int_{s_0}^\infty ds \frac{\rho_2^h(s)}{s-(p+q)^2} +
\mbox{subtractions}\, ,
\end{eqnarray}
where $\rho_{1,2}^h$ show the spectral densities of the higher
resonances and the continuum in hadronic representation. These spectral densities are  approximated by evoking the quark-hadron duality assumption,
\begin{equation}\label{eq26}
\rho_{1,2}^h(s)=\rho_{1,2}^{QCD}(s)\theta (s-s_0),
\end{equation}
where, $\rho_{1,2}^{QCD}(s)=\frac{1}{\pi}Im\Pi^{QCD}(s)$ are spectral densities in QCD side and $s_0$ is continuum threshold in $D_s$ channel.

The  correlation function in QCD side, $\Pi^{QCD}(s)$ is
calculated  by expanding the $T$ product of the currents in
(\ref{eq21}) in terms of the DA's of the $\eta$ meson with
increasing twist in deep Euclidean region, where $(p+q)^2\ll0$.
After contracting out the $c$ quark pair, we obtain
\begin{eqnarray} \label{eq27}
\Pi_\mu (p,q)&=&i \int d^4x e^{iqx} \langle \eta \vert \bar{s}
\gamma_\mu
(1-\gamma_5) S_c(x) (1-\gamma_5) s(0) \vert 0 \rangle, 
\end{eqnarray}
where, $S_c(x)$ is the full propagator of $c$ quark. 

The light cone expansion of the quark propagator in the external  gluon field is
made in \cite{Balitsky}. The propagator receives contributions from higher Fock
states proportional to the condensates of the operators $\bar q Gq$, $\bar q GGq$ and $\bar q q \bar q q$. In the present work, we neglect contributions 
with two gluons as well as four quark operators due to the fact that
their contributions are small \cite{Braunf}. In this approximation, the  $S_c(x)$ is given as:
\begin{eqnarray}\label{eq28}
S_c(x)&=&\int \frac{d^4k}{(2\pi)^4}
e^{-ikx} \frac{\not\!k + m_c}{k^2-m_c^2}-ig_s\int
\frac{d^4k}{(2\pi)^4}e^{-ikx}\int_0^1du\left[\frac{1}{2}\frac{k\!\!\!/+m_c}{(m_c^2-k^2)^2}G_{\mu\nu}(ux)\sigma^{\mu\nu}
\right.\nonumber\\
&+&\left.\frac{1}{m_c^2-k^2}ux_\mu G^{\mu\nu}(ux)\gamma_\nu\right],
\end{eqnarray}
where $G_{\mu\nu}$ is the gluonic field strength tensor and  $g_s$ is the
strong coupling constant. We can rewrite the Eq. (\ref{eq27}) as:
\begin{eqnarray}\label{eq28888}
\Pi_\mu(p,q)&=&\frac{i}{4} \int d^4x e^{iqx} \Big[\mbox{\rm Tr} \gamma_\mu
(1-\gamma_5) S_c(x)(1-\gamma_5) \Gamma_i \Big]\langle \eta \vert
\bar{s} \Gamma^i s \vert 0 \rangle,
\end{eqnarray}
where $\Gamma^i$ is the full set of the Dirac matrices, $\Gamma^i =
(I,~\gamma_5,~\gamma_\alpha,~\gamma_\alpha
\gamma_5,~\sigma_{\alpha\beta})$.
As it is clear from Eq. (\ref{eq28888}), to proceed to
 calculate the theoretical side of the correlation
function, we need to know the matrix elements of the nonlocal operators between vacuum and
$\eta$ meson  states. Up to twist-4, the
$\eta$ meson DA's are defined as
\cite{Ball} :
\begin{eqnarray}
\langle \eta(p) | \bar q \gamma_\mu \gamma_5 q | 0 \rangle &=& -i
f_\eta p_\mu \int_0^1 du e^{-i u p x} \left[
\varphi_\eta (u) + \frac{1}{16} m_\eta^2 x^2 A(u) \right] \nonumber \\
&& - \frac{i}{2} f_\eta m_\eta^2 \frac{x_\mu}{px} \int_0^1 du
e^{-i u px} B(u),
\label{eq211}\\
\langle \eta(p) | \bar q(x) \gamma_\mu \gamma_5 g_s G_{\alpha
\beta}(v x) q(0) | 0 \rangle &=& f_\eta m_\eta^2 \left[ p_\beta
\left( g_{\alpha \mu} - \frac{x_\alpha p_\mu}{px} \right)
-p_\alpha \left( g_{\beta \mu} - \frac{x_\beta p_\mu}{px} \right)
\right] \nonumber \\ && \times \int {\cal D}\alpha_i
\varphi_\perp(\alpha_i) e^{-i p x(\alpha_1 + u \alpha_3)} \nonumber \\
&& + f_\eta m_\eta^2 \frac{p_\mu}{px} (p_\alpha x_\beta - p_\beta
x_\alpha) \int {\cal D}\alpha_i \varphi_\parallel(\alpha_i) e^{-i
p x(\alpha_1 + u \alpha_3)},
\nonumber \\
\label{eq212}\\
\langle \eta(p) | \bar q(x) g_s \tilde G_{\alpha \beta}(vx)
\gamma_\mu q(0) | 0 \rangle &=& i f_\eta m_\eta^2 \left[ p_\beta
\left( g_{\alpha \mu} - \frac{x_\alpha p_\mu}{px} \right)
-p_\alpha \left( g_{\beta \mu} - \frac{x_\beta p_\mu}{px} \right)
\right] \nonumber \\ && \times \int {\cal D}\alpha_i \tilde
\varphi_\perp(\alpha_i) e^{-i p x(\alpha_1 + u \alpha_3)} \nonumber \\
&& +i f_\eta m_\eta^2 \frac{p_\mu}{px} (p_\alpha x_\beta - p_\beta
x_\alpha) \int {\cal D}\alpha_i \tilde \varphi_\parallel(\alpha_i)
e^{-i p x(\alpha_1 + u \alpha_3)},
\nonumber \\
\label{eq213}
\end{eqnarray}
where, $\tilde G_{\mu \nu} = \frac{1}{2} \epsilon_{\mu \nu \sigma
\lambda} G^{\sigma \lambda}$ and ${\cal D} \alpha_i = d \alpha_1 d
\alpha_2 d \alpha_3 \delta(1-\alpha_1 - \alpha_2 - \alpha_3)$.
Since we use the chiral current, the twist-3 wave functions do not give
any contribution. In Eqs. (\ref{eq211})-(\ref{eq213}), the
$\varphi_\eta(u)$ is the leading twist-2, $A(u)$ and part of
$B(u)$ are two particle twist-4, $\varphi_\parallel(\alpha_i)$,
$\varphi_\perp(\alpha_i)$, $\tilde \varphi_\parallel(\alpha_i)$
and $\tilde \varphi_\perp(\alpha_i)$ are three particle twist-4
DA's.
Here we should stress that using the identity,
\begin{eqnarray}
\gamma_\mu \sigma_{\alpha \beta} = i (g_{\mu \alpha} \gamma_\beta
- g_{\mu \beta} \gamma_\alpha ) + \epsilon_{\mu \alpha \beta \rho}
\gamma^\rho \gamma_5\, ,
\end{eqnarray}
 and due to the parity invariance of strong interactions, the matrix element,
\begin{eqnarray}
 \langle \eta(p)|
\bar s \gamma_\mu G^{\alpha \beta}(u x) \sigma_{\alpha \beta} s |
0 \rangle=0,
\end{eqnarray}
and has no contribution.
 For extracting the QCD or theoretical side of the correlation function,  we insert  the expression of the charm quark full propagator as well as the DA's of the $\eta$ meson into Eq. (\ref{eq28888})
and carry out the Fourier transformation.

Now, we proceed to get the LCSR for our form factors equating the
coefficients of the corresponding $p_\mu$ and $q_\mu$ structures
from both  phenomenological and QCD  sides of the correlation function and
applying Borel transform with respect to the variable $(p+q)^2$ in
order to suppress the contributions of the higher states and
continuum as well as eliminate the subtraction terms.
As a result, the following sum rules for the form factors $f_+^{D_s\to\eta}$ and
$f_+^{D_s\to\eta} + f_-^{D_s\to\eta}$ are obtained:
\begin{eqnarray}\label{eq214}
f_+^{D_s\to\eta}(q^2)&=&\frac{m_c^2 m_\eta^2 f_\eta}{2 m_{D_s}^2
f_{D_s}}e^{\frac{m_{D_s}^2}{M^2}} \Bigg\{\int_\delta^1
\frac{du}{u} \left(\frac{2\varphi_\eta(u)}{
m^2_\eta}+\frac{3A(u)}{4uM^2}-\frac{m_c^2A(u)}{2u^2M^4}\right)
e^{\frac{-s(u)}{M^2}}
\nonumber\\
&&+2\int_\delta^1 du \int_0^u dt
\frac{B(u)}{tM^2}e^{\frac{-s(u)}{M^2}}
\nonumber\\
&&-\int_\delta^1 du \int {\cal D}\alpha_i
\frac{8\varphi_\perp(\alpha_i)+2\varphi_\parallel(\alpha_i)
-8\tilde\varphi_\perp(\alpha_i)-2\tilde
\varphi_\parallel(\alpha_i)}{k^2M^2}e^{\frac{-s(k)}{M^2}}
\nonumber\\
&&+4m^2_\eta \int_\delta^1 du \int {\cal D}\alpha_i \int_0^k dt
\frac{\varphi_\perp(\alpha_i)+\varphi_\parallel(\alpha_i)
-2\tilde\varphi_\perp(\alpha_i)-2\tilde
\varphi_\parallel(\alpha_i)}{t^2M^4}e^{\frac{-s(t)}{M^2}}\Bigg\}\, ,
\end{eqnarray}
\begin{eqnarray}
f_+^{D_s\to\eta}(q^2) + f_-^{D_s\to\eta}(q^2)&=&\frac{m_c^2 m_\eta^2
f_\eta}{m_{D_s}^2 f_{D_s}}e^{\frac{m_{D_s}^2}{M^2}}
\Bigg\{2\int_\delta^1 du \int_0^u dt \frac{B(u)}{t^2
M^2}e^{\frac{-s(t)}{M^2}}
\nonumber\\
&&-4m^2_\eta \int_\delta^1 du \int {\cal D}\alpha_i \int_0^kdt
\frac{2\varphi_\perp(\alpha_i)+2\varphi_\parallel(\alpha_i)-\tilde\varphi_\perp(\alpha_i)-\tilde
\varphi_\parallel(\alpha_i)}{t^3M^4}e^{\frac{-s(t)}{M^2}}\Bigg\},\nonumber\\
\end{eqnarray}
where, $M^2$ is the Borel parameter and,
\begin{eqnarray}\label{eq215}
s(x)&=& \frac{m_c^2-q^2\bar{x}+m_\eta^2 x \bar{x}}{x}\, ,\nonumber \\
\bar{x}&=& 1-x\, ,\nonumber \\
k&=&\alpha_1+u\alpha_3\, ,\nonumber \\
\delta &=& \frac{1}{2 m_\eta^2} \Big[(m_\eta^2+q^2-s_0) +
\sqrt{(s_0-m_\eta^2-q^2)^2-4 m_\eta^2(q^2-m_c^2)}\Big]\, .
\end{eqnarray}

\section{Numerical analysis}
In this section, we  numerical analyze  the form factors, $%
f_\pm^{D_s\to(\eta,\eta^{\prime})}(q^2)$  and calculate branching
fractions of the $D_s\to (\eta,\eta^{\prime}) l \nu$ decays and their ratio. We also compare the results of the considered observables with predictions of the other nonperturbative approaches as well as existing experimental data.
 As we
mentioned before, using  Eq. (\ref{eq14}), the transition form
factors of $D_s\to \eta' l \nu$ decay are calculated by the help
of the transition form factors of $D_s\to \eta l \nu$ decay
easily. Hence, we will discuss only the $f_\pm^{D_s\to\eta}(q^2)$
form factors. From the LCSR for these form factors, it follows
that the main input parameters  are the DA's of the $\eta$ meson.
The explicit expressions of the wave functions, $\varphi_\eta(u)$,
$A(u)$, $B(u)$ and $\varphi_\parallel(\alpha_i)$,
$\varphi_\perp(\alpha_i)$, $\tilde \varphi_\parallel(\alpha_i)$,
and $\tilde \varphi_\perp(\alpha_i)$ as well as related parameters
are given as \cite{Ball}:
\begin{eqnarray}
\varphi_{\eta}(u) &=& 6 u \bar u \left( 1 + a_2^{\eta} C_2^{3
\over 2}(2 u - 1) \right),
\nonumber \\
\tilde\varphi_\parallel(\alpha_i) &=& 120 \alpha_1 \alpha_2
\alpha_3 \left( v_{00} + v_{10} (3 \alpha_3 -1) \right),
\nonumber \\
\varphi_\parallel(\alpha_i) &=& 120 \alpha_1 \alpha_2 \alpha_3
\left( a_{10} (\alpha_2 - \alpha_1) \right),
\nonumber\\
\tilde\varphi_\perp (\alpha_i) &=& - 30 \alpha_3^2\left[
h_{00}(1-\alpha_3) + h_{01} (\alpha_3(1-\alpha_3)- 6 \alpha_2
\alpha_1) +
    h_{10}(\alpha_3(1-\alpha_3) - \frac32 (\alpha_1^2+ \alpha_2^2)) \right],
\nonumber\\
\varphi_\perp (\alpha_i) &=& 30 \alpha_3^2(\alpha_1 - \alpha_2)
\left[ h_{00} + h_{01} \alpha_3 + \frac12 h_{10}(5 \alpha_3-3)
\right],
\nonumber \\
B(u)&=& g_{\eta}(u) - \varphi_{\eta}(u),
\nonumber \\
g_{\eta}(u) &=& g_0 C_0^{\frac12}(2 u - 1) + g_2 C_2^{\frac12}(2 u
- 1) + g_4 C_4^{\frac12}(2 u - 1),
\nonumber \\
{A}(u) &=& 6 u \bar u \left[\frac{16}{15} + \frac{24}{35}
a_2^{\eta}+ 20 \eta_3 + \frac{20}{9} \eta_4 +
    \left( - \frac{1}{15}+ \frac{1}{16}- \frac{7}{27}\eta_3 w_3 - \frac{10}{27} \eta_4 \right) C_2^{3 \over 2}(2 u - 1)
    \right. \nonumber \\
    &+& \left. \left( - \frac{11}{210}a_2^{\eta} - \frac{4}{135} \eta_3w_3 \right)C_4^{3 \over 2}(2 u - 1)\right]
\nonumber \\
&+& \left( -\frac{18}{5} a_2^{\eta} + 21 \eta_4 w_4 \right)\left[
2 u^3 (10 - 15 u + 6 u^2) \ln u \right. \nonumber\\ &+& \left. 2
\bar u^3 (10 - 15 \bar u + 6 \bar u ^2) \ln\bar u + u \bar u (2 +
13 u \bar u) \right] \label{wavefns},
\end{eqnarray}
where $C_n^k(x)$ are the Gegenbauer polynomials,
\begin{eqnarray}
h_{00}&=& v_{00} = - \frac13\eta_4,
\nonumber \\
a_{10} &=& \frac{21}{8} \eta_4 w_4 - \frac{9}{20} a_2^{\eta},
\nonumber \\
v_{10} &=& \frac{21}{8} \eta_4 w_4,
\nonumber \\
h_{01} &=& \frac74  \eta_4 w_4  - \frac{3}{20} a_2^{\eta},
\nonumber \\
h_{10} &=& \frac74 \eta_4 w_4 + \frac{3}{20} a_2^{\eta},
\nonumber \\
g_0 &=& 1,
\nonumber \\
g_2 &=& 1 + \frac{18}{7} a_2^{\eta} + 60 \eta_3  + \frac{20}{3}
\eta_4,
\nonumber \\
g_4 &=&  - \frac{9}{28} a_2^{\eta} - 6 \eta_3 w_3 \label{param0}.
\end{eqnarray}
The constants in the Eqs.~(\ref{wavefns}) and (\ref{param0}) were
calculated at the renormalization scale $\mu=1 ~~GeV^{2}$ using
QCD sum rules  and are given as $a_{2}^{\eta} = 0.2$,
$\eta_{3} =0.013$, $\eta_{4}=0.5$, $w_{3} = -3$ and $ w_{4}= 0.2$.

The values of the other input parameters appearing in sum rules
for form factors are: qauark masses at
the scale of about $1~GeV$ 
 $m_s = 0.14~GeV$, $m_c=1.3~GeV$ \cite{B.L.Ioffe.bey}, meson masses
$m_{\eta}=0.5478~GeV$, $m_{\eta'}=0.9578~GeV$,
$m_{D_s}=1.9685~GeV$, $V_{cs}=1.023\pm 0.036 $ \cite{PDG2010} and
$f_{D_s}=(0.274\pm0.013\pm0.007)~GeV$ \cite{Artu}.

\begin{figure}[th]
\begin{center}
\begin{picture}(160,30)
\put(0,-20){ \epsfxsize=8cm \epsfbox{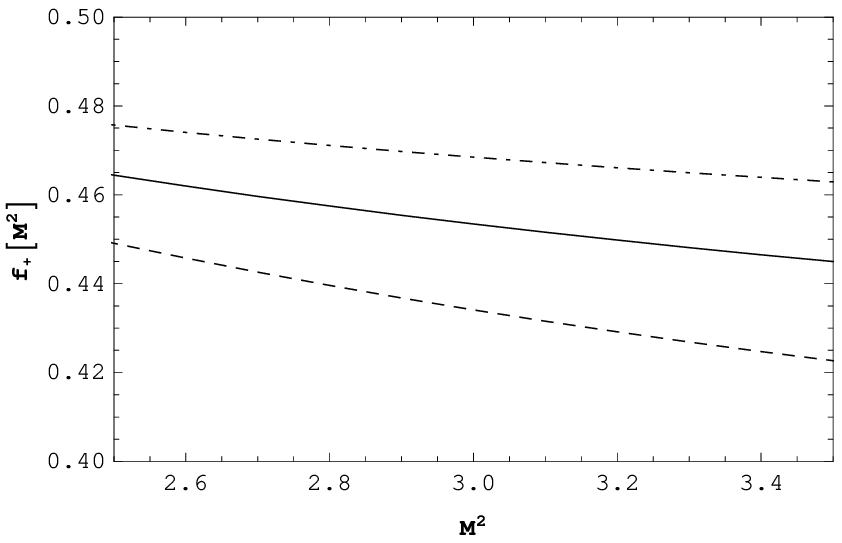}\epsfxsize=8cm
\epsfbox{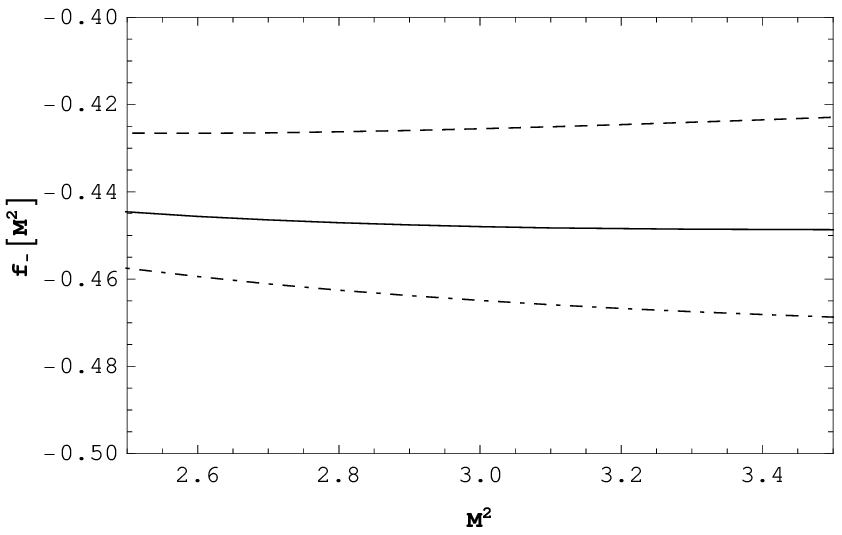} }
\end{picture}
\end{center}
\vspace*{1cm} \caption{The dependence of the form factors on $M^2$. The dashed, solid  and dashed-dotted lines correspond to the
 $s_0=5.5$, $s_0=6$ and $s_0=6.5$, respectively.}\label{F31}
\end{figure}

The sum rules for form factors also contain two auxiliary
parameters, $s_0$ and $M^2$.   The continuum threshold is not
totally arbitrary but it depends on the energy of the first
excited state. We choose, $s_0=(6.5 \pm 0.5)~GeV^2$ (see also
\cite{Colozp}). Now, we are looking for a working region for
$M^2$, where  according to sum rules philosophy, our  numerical
results be stable for a given continuum threshold $s_0$. The
working region for the Borel mass parameter is determined
requiring that not only  contributions of the higher states and
continuum effectively suppress,
 but also contributions of the DA's with higher twists are small. Our numerical analysis shows
that the suitable region is: $2.5~GeV^2 \le M^2 \le 3.5~GeV^2$.
The dependence of the form factors $f_+^{D_s\to\eta}$ and
$f_-^{D_s\to\eta}$ on $M^2$ are shown in Fig. \ref{F31}.
This figure  shows that  the form factors weakly depend on  the Borel mass parameter in its  working region.

Now, we proceed to find the $q^2$ dependence of the form factors. It should be stressed that in the region, $q^2 \ge 1.4~GeV^2$
the applicability of the LCSR is problematic. In order to extend
our results to the whole physical region, we look for a
parametrization of the form factors  such  that in the
region, $0 \le q^2 \le 1.4 ~GeV^2$, the results obtained from the above--mentioned
parametrization coincide well with the light cone QCD sum rules
predictions. The most simple parametrization of the $q^2$
dependence of the form factors is expressed in terms of three
parameters in the following form:
\begin{equation}  \label{eq31}
f_{\pm}(q^2)=\frac{f_{\pm}(0)}{1- \alpha\hat{q}+ \beta\hat{q}%
^2}~,
\end{equation}
where, $\hat{q}=q^2/m_{D_s}^2$.
The values of the parameters, $%
f_\pm^{D_s\to\eta}(0), ~\alpha$ and  $\beta$ are given in  Table
\ref{T31}.  This Table also contains
predictions of the light-front quark model (LFQM) for $f_+^{D_s\to\eta}(0)$ for two sets (for details see \cite{Weike}). The errors presented in this Table are due to
variation of the continuum threshold $s_0$,  variation of the
Borel parameter $M^2$, and uncertainties coming from the DA's and other input parameters.
\begin{table}[th]
\caption{Parameters appearing in the fit function for form factors
of $D_s \to \eta$ in two approaches.}\label{T31}
\begin{ruledtabular}
\begin{tabular}{cccc}
 Model               &$f_-^{D_s\to\eta}(0)$&$\alpha       $&$\beta       $   \\ \hline
 This Work(LCSR)           &$-0.44\pm0.13     $&$2.05\pm0.65  $&$1.08\pm0.35 $   \\ \hline\hline
                     &$f_+^{D_s\to\eta}(0)$&$\alpha       $&$\beta       $   \\ \hline
 This Work (LCSR)          &$0.45\pm0.14     $&$1.96\pm0.63  $&$1.12\pm0.36 $   \\
 LFQM(I)\cite{Weike} &$0.50            $&$1.17         $&$0.34        $   \\
 LFQM(II)\cite{Weike}&$0.48            $&$1.11         $&$0.25        $
\end{tabular}
\end{ruledtabular}
\end{table}

\begin{figure}[th]
\begin{center}
\begin{picture}(160,30)
\put(0,-20){ \epsfxsize=8cm \epsfbox{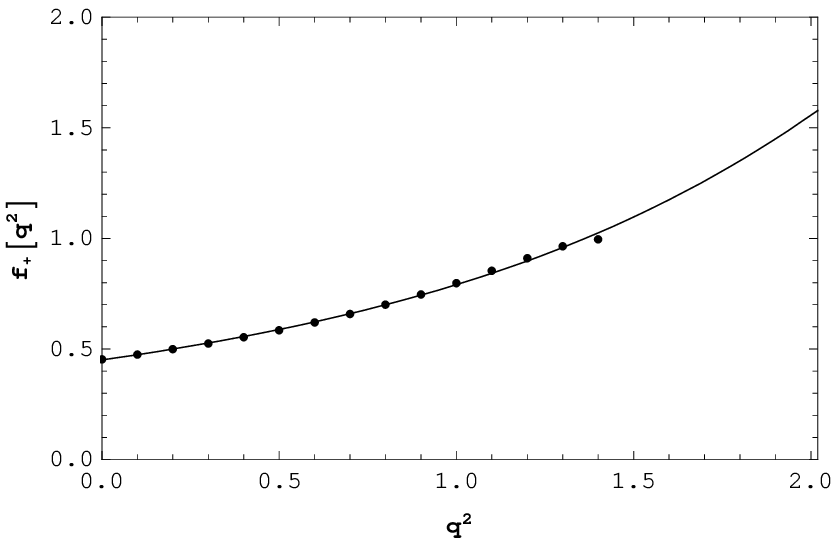}\epsfxsize=8cm
\epsfbox{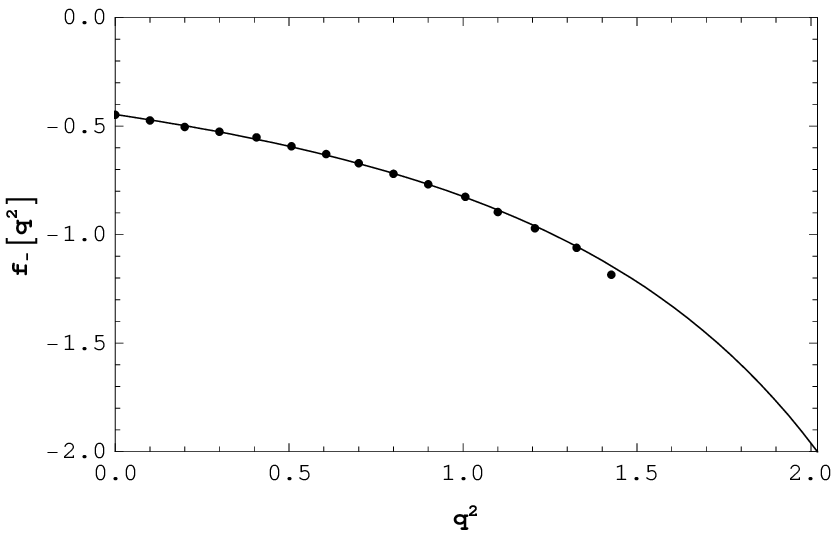} }
\end{picture}
\end{center}
\vspace*{1cm} \caption{The dependence of the form factors of $D_s\to\eta$ on $q^2$.
The circle points correspond to the values obtained directly from
sum rules and  the solid lines belong to the fit parametrization of
the form factors. }\label{F32}
\end{figure}

The dependence of the form factors, $f_{+}(q^2)$ and  $f_{-}(q^2)$ for $D_s\to\eta$
on $q^2$ extracted from the fit parametrization  are shown in Fig.
(\ref{F32}). This figure also contains the form factors obtained directly from our sum rules in reliable
region. We see that, the aforementioned fit parametrization
describe our form factors well.
The values
of the $f_+^{D_s\to(\eta,\eta^{\prime})}(q^2)$ form factors  at $q^2=0$ extracted from fit parametrization and using Eq. (\ref{eq14}) are
shown in Table \ref{T32}. Note that for massless leptons, the form factors, $f_-^{D_s\to(\eta,\eta^{\prime})}(q^2)$ do not contribute to the decay rate formula, so we present only the $f_+^{D_s\to(\eta,\eta^{\prime})}(q^2)$
in this Table. For comparison, the predictions of the
other approaches are also presented in this Table. From this Table, we see a good consistency among the results predicted by different approaches.
\begin{table}[h]
\caption{The $f_+^{D_s\to(\eta,\eta^{\prime})}(q^2)$ form factors at $q^2=0$ in different approaches: this work (LCSR),
three-point QCD sum rules (3PSR) and LFQM. Our results for $f_+^{D_s\to\eta^{\prime}}$ correspond to
$\varphi=39.7^\circ(41.5^\circ)$.}\label{T32}
\begin{ruledtabular}
\begin{tabular}{ccccc}
              Form factor & This work (LCSR) & 3PSR\cite{Colfaz} & LFQM(I)\cite{Weike} & LFQM(II)\cite{Weike}\\
\hline
             $f_+^{D_s\to\eta}(0)$& $0.45\pm0.14  $&$0.50\pm0.04$&$0.50$&$0.48$\\
             $f_+^{D_s\to\eta^{\prime}}(0)$& $0.55\pm0.18
             (0.51\pm0.16)$&$-$&$0.62$&$0.60$
\end{tabular}
\end{ruledtabular}
\end{table}

Now, we would like to evaluate  the branching ratios for the
considered decays. Using the parametrization of the transition matrix elements
in terms of  form factors, in massless lepton case, we get:
\begin{equation}\label{eq32}
\frac{d\Gamma}{dq^2}(D_s \to (\eta,\eta^{\prime}) l \nu_l) =
\frac{G_F^2 |V_{cs}|^2}{
192\pi^3m_{D_s}^3}\left[(m_{D_s}^2+m_{\eta^{(\prime)}}^2-q^2)^2-4m_{D_s}^2m_{(\eta,\eta^{\prime})}^2\right]^{3/2}
|f_+^{D_s\to\eta^{(\prime)}}(q^2)|^2 \, ,
\end{equation}
where $G_F$ is the Fermi constant. Integrating Eq. (\ref{eq32})
over $q^2$ in the whole physical region and using the total mean
lifetime, $\tau_{D_{s}}= (0.5\pm0.007)~ps$ \cite{PDG2010}, the
branching ratios of the $D_s \to (\eta,\eta^{\prime})l \nu$ decays
are obtained as presented in Table \ref{T33}.
\begin{table}[th]
\caption{The branching ratios in different models and experiment. Our values correspond to $39.7^\circ(41.5^\circ)$.}
\label{T33}
\begin{ruledtabular}
\begin{tabular}{c|ccccc}
Mode&This work&3PSP\cite{Colfaz}&LFQM(I)\cite{Weike}&LFQM(II)\cite{Weike}&EXP\cite{PDG2010}\\
\hline
Br$(D_{s} \to \eta l \nu)\times 10^{2}$& $3.15\pm0.97 $             & $2.3\pm0.4$ & $2.42$ & $2.25$&$2.9\pm0.6  $ \\
Br$(D_{s} \to \eta'l \nu)\times 10^{2}$& $0.97\pm0.38(0.84\pm0.34)$ & $1.0\pm0.2$ & $0.95$ & $0.91$&$1.02\pm0.33$  \\
\end{tabular}
\end{ruledtabular}
\end{table}
This Table  also includes a comparison of our results and
predictions of the other nonperturbative approaches including the
LFQM  and 3PSR
 and experimental values \cite{PDG2010}. From this Table,
we see a good consistency between our results and predictions of
the  different approaches especially  experimental data.

At the end of this section, we would like to compare also the
ratio: $\mbox{R}_{D_s}=\frac{\mbox{Br}(D_s\to \eta' l
\nu)}{\mbox{Br}(D_s\to \eta l \nu)}$ in Table \ref{T34} for
different approaches as well as  experimental value. This Table
also depicts a good consistency among the values, specially
between our prediction with $\varphi=39.7^\circ$ and experimental
value. This can be considered as a good test for correctness of
the
 considered internal structure for the $D_s$ meson as well as the mixing angle between  $\eta$ and $\eta'$ states.
\begin{table}[th]
\caption{The $\mbox{R}_{D_s}$  with respect to mixing angle, $\varphi$
 for different models and experimental value.} \label{T34}
\begin{ruledtabular}
\begin{tabular}{ccc}
Model & Angle $(\varphi^\circ)$ & $\mbox{R}_{D_s}$\\
\hline
This work (LCSR)           & $39.7^\circ(41.5^\circ)$ & $0.32\pm0.02(0.27\pm0.01)$ \\
3PSR\cite{Colfaz}   & $40^\circ              $ & $0.44\pm0.01             $ \\
LFQM(I)\cite{Weike} & $39^\circ              $ & $0.39                    $ \\
LFQM(II)\cite{Weike}& $39^\circ              $ & $0.41                    $ \\
EXP\cite{PDG2010}         & $-                     $ &
$0.35\pm0.12$
\end{tabular}
\end{ruledtabular}
\end{table}

\section*{Acknowledgments}
Partial support of Shiraz university research council is
appreciated.

\end {document}